# 2.6mJ/100Hz CEP stable near-single-cycle 4μm laser based on OPCPA and hollow-core-fiber compression

Pengfei Wang,[1,2] Yanyan Li,[1] Wenkai Li,[1,2] Hongpeng Su,[1,2] Beijie Shao,[1,2] Shuai Li, [1,2] Wang Cheng,[1] Ding Wang,[1] Ruirui Zhao,[1,2] Yujie Peng,[1,*] Yuxin Leng,[1,*] Ruxin Li,[1,*] Zhizhan Xu[1]

[1] State Key Laboratory of High Field Physics, Shanghai Institute of Optics and Fine Mechanics, Chinese Academy of Sciences, 390# Qinghe Road, Jiading District, Shanghai 201800, China

[2] University of Chinese Academy of Science, Beijing 100190, China

*Corresponding author: yjpeng@siom.ac.cn, lengyuxin@mail.siom.ac.cn, ruxinli@mail.shcnc.ac.cn

**Abstract**: A carrier envelope phase stable near-single cycle mid-infrared laser based on optical parametric chirped pulse amplification and hollow-core-fiber compression is demonstrated. 4 μm laser pulses with 11.8 mJ energy are delivered from a KTA based OPCPA with 100 Hz repetition rate, and compressed to be ~105 fs by a two-grating compressor with efficiency over 50%. Subsequently, the pulse spectrum is broadened by employing a krypton gas-filled hollow-core-fiber (HCF). Then, the pulse duration is further compressed to 21.5 fs through a $CaF_2$ bulk material with energy of 2.6 mJ and stability of 0.9% RMS, which is about 1.6 cycle for 4 μm laser pulse. The near-single cycle 4 μm laser pulse CEP is passively stabilized with ~370 mrad based on a CEP stable 4 μm OPA injection.

Recently high energy carrier envelope phase (CEP) stable near-single-cycle pulses in the mid-infrared (MIR) from 3-5 μm spectral range attracted much attention in many areas of unexplored strong field physics, such as high harmonic generation (HHG) for extending cutoff energy to keV range to generate broadband and single ultrashort coherent soft x-ray pulse [1-3]. Further, higher repetition rate is required for higher flux of coherent soft x-ray pulses. Tens and hundreds of millijoule MIR pulses with sub-100fs pulse duration have been obtained [4-6] with optical parametric amplification (OPA) and optical parametric chirped pulse amplification (OPCPA) based on suitable nonlinear crystals. Furthermore, in order to generate few-cycle MIR pulses, ultra-broadband parametric amplifications with bandwidth spanning nearly one octave have been developed, e.g., frequency-domain optical parametric amplification (FOPA) [7], dual-chirped OPA (DC-OPA) [8], pulse synthesizer from two-color or multi-color pulses [9,10], and quasi-phase-matched OPA based on periodically poled crystals [11]. With these methods, laser pulses at 1.8μm to 4.2μm with bandwidth of several hundred nanometers and pulse duration of shorter than 2 optical-cycle have been obtained. More importantly, several robust and popular schemes of the pulse self-compression and post-compression techniques have been proposed to reduce the pulse duration to few optical cycles. Both the two further compression processes involve external pulse nonlinear-optical spectral broadening with self-phase modulation (SPM) in bulk materials, air or noble gas filled capillaries firstly, and dispersion compensation with chirped mirrors or bulk materials secondly [12,13], which are widely used to generate millijoule few-cycle pulses in MIR range. B. E. Schmidt et al. demonstrated this scheme to generate sub-millijoule level few-cycle duration experimentally and theoretically at 1.8μm [14, 15], and A. Baltuska et al. have done numerous studies to compress 5mJ/ 80fs down to 2.5mJ/22fs at

3.2μm [16] and 35mJ/80fs down to 35fs at 3.9μm [17] successfully.

In this letter, we present a CEP stable near-single-cycle 4 μm laser system based on a collinear OPCPA and HCF further compression, which can deliver laser pulses with 2.6 mJ/21.5 fs operating at 100 Hz repetition rate. The spectral broadening evolution process has been investigated by changing the noble gas pressure filled in the HCF. With suitable chirp and energy of incident pulses and appropriate gas pressure in the hollow core fiber (HCF), near-single-cycle pulses can be obtained. Finally, the CEP is passively stabilized with 370 mrad (RMS) benefit from an injection of a femtosecond OPA idler.

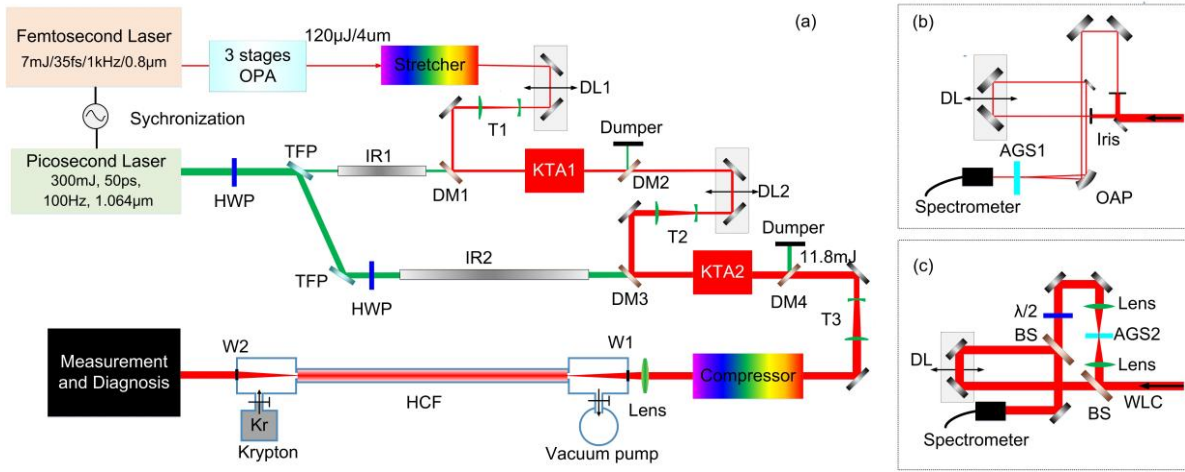

**Fig. 1.** (a) Schematic of the 4 μm OPCPA and post-compression system. (b) The second-harmonic-generation-frequency-resolved optical gating (SHG-FROG) measurement. (c) The $f$ - $2f$ equipment for CEP measurement. DL, delay line; HWP, half wave plate; TFP, thin film polarizing; DM, dichroic mirror; T, telescope; W, window made of CaF2; OAP, off-axis parabolic mirror; WLC, white light output from HCF; AGS, AgGaS2 crystal; BS, beam splitter.

The schematic of the experimental setup is presented in Fig. 1(a). The original CEP stable 4 um laser pulses with about 120 uJ energy are from a home-built OPA device pumped by a commercial Ti:sapphire femtosecond laser (Coherent Inc.) [18]. A conventional Offnër-type stretcher with ~20% efficiency is used to stretch the pulse duration to 50 ps as the signal light injection of the following OPCPA amplifier. The pump laser for the OPCPA amplifier is also from a home-built picosecond Nd:YAG laser which can deliver 1064 nm laser pulses with up to 300 mJ energy and 50 ps pulse duration running at 100 Hz [19]. An electric synchronization system is employed to lock the same repetition rate of both the oscillators of the commercial Ti:sapphire laser and pump laser at 80 MHz. The pump beam is split and image relayed to the nonlinear crystals with 50 mJ/Φ3.5 mm for the first stage amplifier and 210 mJ/Φ6 mm for the second stage amplifier respectively. Two potassium titanyle srsenate crystals, $KTiOAsO_4$ (KTA), are used in the parametric amplifiers for their good transparency and nonlinear optical properties. Both KTA crystals are cut at $\theta$=40.8 ° for type II phase matching with 10 mm thickness. The amplified chirped pulses with 11.8 mJ energy are compressed to 105 fs by employing a two grating compressor. In order to obtain near-single-cycle pulses at 4 μm, a further compressor based on HCF is used. A HCF with 1 mm core diameter and 3 meters length is chosen to obtain a small propagate losses and large nonlinear SPM simultaneously, which is provided by Few-Cycle Inc. The pressure of the krypton gas filled in HCF is controlled precisely and monitored in real-time. The CaF2 optics used in the HCF system can compensate the chirp caused by the SPM process. The characterizations of the final output pulses are measured and diagnosed by a home-built SHG-FROG and an $f$ - $2f$ measurements as shown in Fig. 1(b) and Fig. 1(c).

By the way, the similar structures and the identical crystals are used in two OPCPA stages to ensure perfect matching of the gain bandwidths and absent the spectral narrowing effects. With 50 mJ and 200 mJ pump energy for the 1st and 2nd KTA crystal, the 4 μm signal pulses are amplified to 1.5 mJ and 11.8 mJ respectively. In Fig. 2, the results of the characterizations of the OPCPA are presented. Fig. 2(a) describes the gain in the 2nd stage high energy OPA as a function of the signal and the pump energies. One can see almost the linear progression of the output amplified energy with the increasing of the pump energy. The spectrum evolution throughout the OPCPA process is shown in Fig. 2(b), and it can be concluded that spectrums of every stage OPA amplifier progress at 4μm are hardly any change at all, and the output spectrum can support a Fourier-transform limit (FTL) duration of 101 fs. Fig. 2(c) and (d) show the amplified beam profile from KTA1 and KTA2, which are near Gaussian distribution with little distortion.

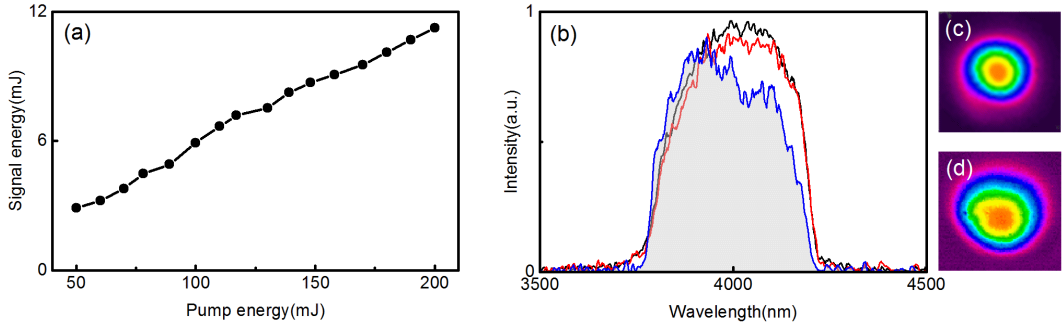

**Fig. 2.** (a) Amplified pulse energy of the 2nd stage OPA as a function of pump energy. (b) The spectrum evolution throughout the OPCPA system. Amplified beam profile after 1st (c) and 2nd (d) OPA.

After the two KTA amplifiers, about 10 mJ energy is delivered to the twograting compressor. By adjusting the compressor precisely, the obtained shortest duration is 105 fs measured with the SHG-FROG measurement as shown in Fig. 3(a-d), and pulse energy of 5.5 mJ is obtained with a total compression efficiency of 55%. Subsequently, the pulses are injected to a krypton gas filled HCF for spectral broadening and further compression. Two precise mounts at the both ends of the HCF are specially designed to keep the fiber straight over a long length to reduce the bending losses and allow pumping and air inflating. This HCF compression system is suited to compress high energy and peak power pulses at mid-IR wavelength range because of the self-focusing critical power scaling with wavelength as $P_{crir} \propto \lambda^2$.

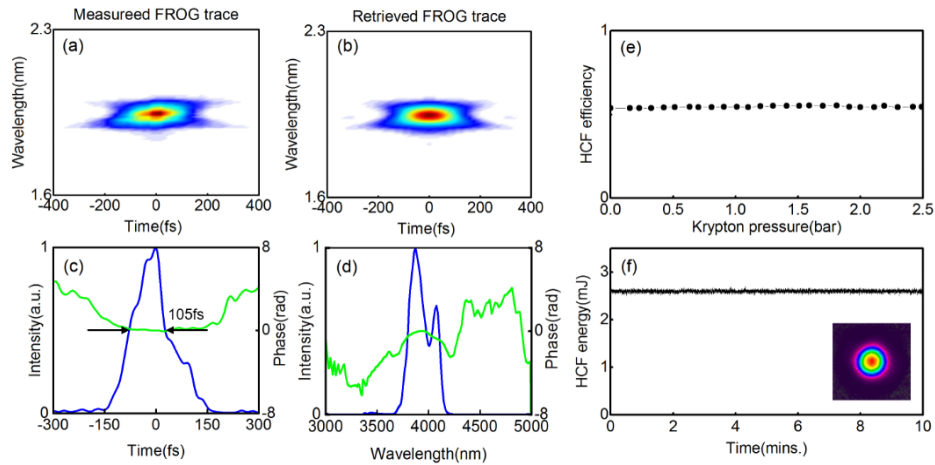

**Fig. 3.** (a) Measured and (b) reconstructed SHG-FROG traces, (c) pulse temporal profile and (d) reconstructed spectrum. (e) Transmission efficiency of HCF system with increasing pressure in HCF. (f) Energy influence of 4 μm pulses with 21.5 fs from HCF based post-compression system.

The optimal focused spot diameter at the injection end of HCF is 0.63 mm by using a f=700mm uncoated CaF2 lens with 90% transmission efficiency, and the input and output windows are placed far enough from the fiber (here is about 500 mm) to avoid unwanted SPM in windows and protect them from high energy of focused beam. With the input pulses energy of 5.0 mJ before HCF input window, about 2.6mJ pulses energy are obtained through output window, corresponding a total transmission of 52%. The throughout efficiency during changing the krypton pressure filled in the HCF from 0 to 2.5 bar is always exceeding 50% as shown in Fig. 3(e). It can be inferred that the pulse compression process does not involve any noticeable ionization of up to 2.5 bar of krypton pressure in HCF. The energy influence of output pulse is about 0.9% (RMS) measured in 10 mins. An advantage of pulse post-compression in HCF is that the output beam quality is much better than the injection owing to the mode selection effect of waveguides, as shown inset in Fig. 3(f).

To get stronger SPM effect, the pulse duration is controlled at 140 fs with a little positive chirp by adjusting the relative distance of the compressor. The SPM effect between input pulses and noble gas atoms contributes to broaden the spectrum of laser pulses. Then, the spectral broadened pulses are recompressed in a 2 mm thick CaF2 plate at the output window. A silver concave mirror is used to collimate the output beam for measurements.

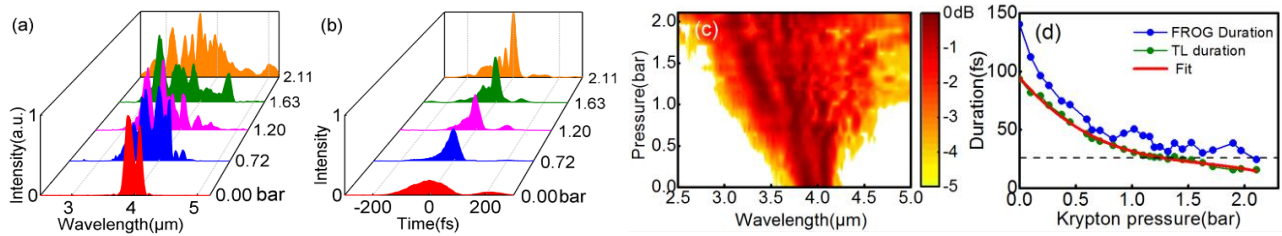

**Fig. 4.** (a) Spectrum and (b) the corresponding temporal profile measured by SHG-FROG at different krypton pressures. (c) Spectrum evolution throughout increasing gas pressures in HCF. (d) Pulse durations from SHG-FROG and Fourier transform limited durations of the spectrally broadened pulses for different krypton pressures, and the dash line is 26 fs, corresponding to two cycles for 4 μm laser pulses.

As the krypton pressure increasing, spectral broadening is more significant. For instance, at 2.0 bar, spectrum after output window span from 2.5 μm to 5.2 μm which is exceeding an octave. The pulse duration is also become narrower simultaneously. The measured spectral and corresponding temporal evolutions of 4 μm pulses throughout increasing the krypton pressures are summarized in Fig. 4. It shows the krypton gas at different pressures in hollow-core-fiber provides a significant spectral broadening due to SPM effect. With the krypton pressure increasing, the spectral broaden are wider, the pulse duration are shorter and the temporal profile get worse. Due to the short-pulse propagation in a long optical fiber, self-steepening would play a significant role and lead to spectrally broadened asymmetric, and shorter wavelength part of spectrum is more easily broaden than longer wavelength, as shown in Fig. 4(a)-(b). And it also brings about a deep pulse trailing edge in temporal profile. When the krypton pressure exceed 1.5 bar, a slight sub-pulse emerging and increasing with higher krypton pressure. The actual measured pulse durations and Fourier transform limited durations (FTL) corresponding to the broadened spectrums at different pressures are shown in Fig. 4(d). It shows that the broadened bandwidth with high krypton pressure is broad enough to support single optical cycle pulse generation.

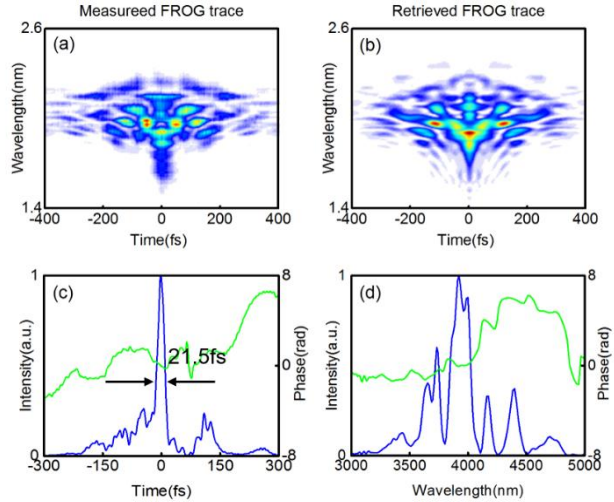

**Fig. 5.** Temporal characterizations of the pulse compression. (a) Measured and (b) reconstructed SHG-FROG traces; (c) pulse temporal profile of 21.5 fs FWHM duration and (d) reconstructed spectrum.

After a series of parameters adjustment on gas pressure of krypton and input pulse energy, it is found that the optimal krypton pressure and input energy for pulse compression and dispersion compensation in a 2 mm-thick window $CaF_2$ is 1.15 bar and 5.5 mJ. Temporal and spectrum characterizations of post-compressed pulses and corresponding spectrum and group delay dispersion (GDD) retrieved from SHG-FROG measurement are shown in Fig. 5. It indicates the pulse duration of 21.5 fs is achieved, corresponding to 1.6 optical cycles and 1.08 FTL (FTL=20 fs). With the krypton pressure increasing, a shorter compressed pulse duration 15 fs could be achieved but there is an obvious pedestal in the temporal profile.

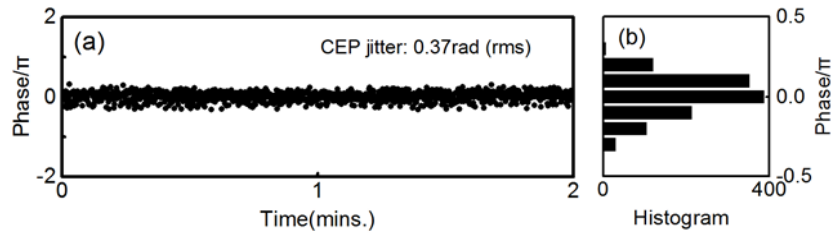

**Fig. 6.** (a) CEP fluctuation measured by $f$ - $2f$ setup. (b) Histogram of CEP.

Benefit from the CEP stable injection from an OPA idler, the amplified and compressed 4 μm pulses are CEP stable to a certain extent. The broadened output spectrum with more than one-octave by increasing the gas pressure is generated for the CEP measurement, instead of white light continuum (WLC) generated in bulk materials. The frequency-doubled spectrum in CEP measure system is obtained by a 0.5-mm thick AGS crystal and interference spectrum is measured by a NIR-QUESR spectrometer. The CEP is calculated by Fourier transform spectral interferometric (FTSI) from every interference spectrum [20], and the result suggests that the CEP stability is 370 mrad in 2 minutes (shown in Fig. 6). This measurement confirms that the CEP stability is preserved during OPCPA and SPM broadening process.

It's worth mentioning that there are two limitations for the output capacity of the laser system. The first one is the damage threshold of coating films of the MIR optics. To keep the safety of the system, only 200 mJ pump energy is used to pump the 2[nd] stage amplifier and the OPA has not been saturation state. The second one is the low transmission efficiency of some optics used in the system, e.g., 4 μm gratings, uncoated lenses. These optical elements cause a terrible efficiency of the whole OPCPA and post-compression system. Therefore, several methods to improve output capacity would be effective such as replacing optics with higher damage threshold and higher transmission efficiency.

In conclusion, we demonstrate a near-single-cycle CEP-stable 4 μm laser system running at 100 Hz, which can generate 2.6 mJ, 21.5 fs (1.6 optical cycles) laser pulses with CEP jitter of 370 mrad. In this laser, OPCPA is used for energy upgrade which has tremendous potential for higher output capability with higher pump, and HCF based further compression is employed which can be applied for MIR laser with wavelength up to 6 μm. The CEP stable 4 μm laser system will be used for high-flux single ultrashort coherent soft x-ray pulse generation by driving HHG.

**Funding**. The Strategic Priority Research Program of the Chinese Academy of Sciences, Grant No. XDB1603; International S&T Cooperation Program of China, Grant No. 2016YFE0119300.